\documentclass[11pt]{article}

\usepackage{acl}

\usepackage{times}
\usepackage{latexsym}

\usepackage[T1]{fontenc}

\usepackage[utf8]{inputenc}

\usepackage{microtype}

\usepackage{inconsolata}

\usepackage{graphicx}

\usepackage{amsmath}
\usepackage{multirow}
\usepackage{booktabs}
\usepackage{balance}
\title{LLM-ForcedAligner: A Non-Autoregressive and Accurate LLM-Based Forced Aligner for Multilingual and Long-Form Speech}

\author{Bingshen Mu\textsuperscript{1}, Xian Shi\textsuperscript{2}, Xiong Wang\textsuperscript{2}, Hexin Liu\textsuperscript{3}, Jin Xu\textsuperscript{2}\footnotemark[1], Lei Xie\textsuperscript{1}\thanks{Corresponding Author.}\\
        \textsuperscript{\rm 1}Audio, Speech and Language Processing Group (ASLP@NPU), School of Computer Science,\\ Northwestern Polytechnical University, Xi’an, China\\
    \textsuperscript{\rm 2}Independent Researcher\\
    \textsuperscript{\rm 3}College of Computing and Data Science, Nanyang Technological University, Singapore
    }

\begin{document}
\maketitle
\begin{abstract}
Forced alignment (FA) predicts start and end timestamps for words or characters in speech, but existing methods are language-specific and prone to cumulative temporal shifts. The multilingual speech understanding and long-sequence processing abilities of speech large language models (SLLMs) make them promising for FA in multilingual, crosslingual, and long-form speech settings. However, directly applying the next-token prediction paradigm of SLLMs to FA results in hallucinations and slow inference. To bridge the gap, we propose \textbf{LLM-ForcedAligner}, reformulating FA as a slot-filling paradigm: timestamps are treated as discrete indices, and special timestamp tokens are inserted as slots into the transcript. Conditioned on the speech embeddings and the transcript with slots, the SLLM directly predicts the time indices at slots. During training, causal attention masking with non-shifted input and label sequences allows each slot to predict its own timestamp index based on itself and preceding context, with loss computed only at slot positions. Dynamic slot insertion enables FA at arbitrary positions. Moreover, non-autoregressive inference is supported, avoiding hallucinations and improving speed. Experiments across multilingual, crosslingual, and long-form speech scenarios show that LLM-ForcedAligner achieves a 69\%\textasciitilde78\% relative reduction in accumulated averaging shift compared with prior methods.
Checkpoint and inference code are available at \url{https://github.com/QwenLM/Qwen3-ASR}.
\end{abstract}

\section{Introduction}
In speech processing, the objective of forced alignment (FA) is to estimate the start and end timestamps of each word or character in a speech signal, given its corresponding transcript.
FA is indispensable in numerous applications, including large-scale speech corpus construction and cleaning, automatic subtitling and word-level highlighting, as well as duration modeling and prosody analysis in speech synthesis.
With the continuous advancement of multilingual and multimodal applications, efficient and accurate FA has become increasingly essential~\citep{tseng2021mandarin, Liu2025align}.

\begin{figure}[t]
\centering
\includegraphics[width=1.0\columnwidth]{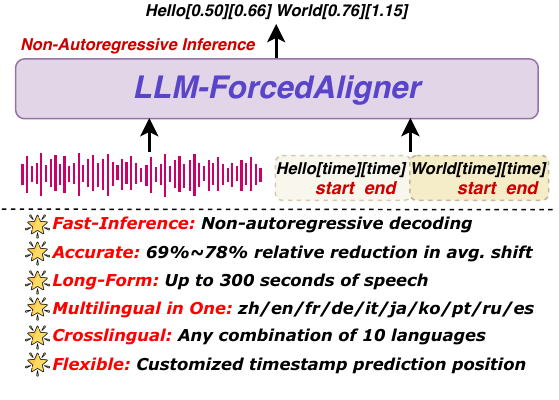}
\caption{Highlights of LLM-ForceAligner.}
\vspace{-22pt}
\label{fig1}
\end{figure}
Existing FA methods can be broadly categorized into two main groups: traditional hybrid systems~\citep{Michael2017mfa} and end-to-end models~\citep{Elena2023nfa, Ludwig2020ctc, xian2023achieving, Max2023whisperx}.
Montreal forced aligner (MFA) is typically a hybrid Gaussian Mixture Model–Hidden Markov Model (GMM–HMM) framework, computing the timestamps via Viterbi decoding for frame-level phoneme-to-text alignment paths~\citep{Michael2017mfa}. Connectionist temporal classification (CTC) is a common end-to-end FA method that leverages frame-to-token alignment computed by CTC-based automatic speech recognition (ASR) models, employing dynamic programming to identify the optimal path that aligns with the text sequence within constrained search paths~\citep{Ludwig2020ctc, Elena2023nfa}.
Continuous integrate-and-fire (CIF) predicts a weight for each encoder output frame and integrates these weights over time. When the accumulated weight exceeds a threshold, a fire event is triggered, at which point a weighted sum of the accumulated frame-level acoustic vectors is computed to generate an acoustic embedding aligned with an output token, thereby enabling the assignment of a corresponding timestamp to each token~\citep{xian2023achieving}.
WhisperX employs a lightweight end-to-end phoneme recognition model to perform frame-level phoneme classification on speech, and then aligns the resulting phoneme sequence with the transcript using dynamic time warping (DTW), thereby obtaining word-level timestamps by aggregating phoneme-level timestamps~\citep{Max2023whisperx}.

However, the aforementioned FA methods are tied to language-specific phonemes, lexicons, or structural designs, which means that in multilingual scenarios, deployment typically involves a collection of independent systems with disparate structures, leading to engineering costs and maintenance complexity that grow linearly with the number of languages.
Furthermore, previous FA methods can be summarized as a process of calculating local acoustic similarities followed by a monotonic path search. While these methods can produce reasonably accurate boundaries for short segments, they frequently accumulate significant systematic temporal shifts in long-form speech.

Large language models (LLMs) have demonstrated powerful abilities in multilingual text understanding and long-sequence processing tasks~\citep{hugo2023llama, hugo2023llama2, an2024qwen25, an2025qwen3}, offering a new possibility for FA that supports multilingual, crosslingual, and long-form speech.
Increasing studies have explored integrating speech encoders with LLMs to build Speech LLMs (SLLMs) that process speech and text within a unified framework.
Nevertheless, existing SLLMs have mainly achieved success in high-level semantic tasks such as ASR~\citep{xuelong2024unveiling, bingshen2025hearing}, speech understanding~\citep{xuelong2025osum, yunfei2023qwenaudio, yunfei2024qwen2audio}, speech synthesis~\citep{xinsheng2025spark, zhihao2025cosyvoice}, and spoken dialogue~\citep{xiong2025freeze, xuelong2025osumechat}. For FA, which is more sensitive to acoustic characteristics, these SLLMs typically treat it as a by-product of ASR and generate word-level or character-level timestamps via next-token prediction. Such a paradigm is susceptible to temporal non-monotonic hallucinations and incurs substantial inference latency.

In this work, we propose a new FA framework named \textbf{LLM-ForcedAligner} that reformulates FA as a slot-filling paradigm: the start and end timestamps of each word or character are treated as discrete time indices, and dedicated special tokens are inserted into the transcript as slots so that, conditioned on the speech embeddings and the transcript augmented with these slots, the SLLM can directly predict the corresponding time indices at the designated slots.
This new FA paradigm effectively leverages the LLM’s strengths in slot filling and long-context processing, extends conventional purely acoustic, phoneme-level alignment to semantic-boundary-aware, character-level or word-level alignment.
To perform slot filling, we apply causal attention masking during training without introducing any shift between the input and label sequences, allowing each slot to predict its own time index based on itself and the preceding context, and we compute the loss function only at the slot positions.
In addition, we adopt a dynamic slot insertion strategy that randomly decides whether to insert special tokens for each word or character in the transcript, enabling LLM-ForcedAligner to predict timestamps for arbitrary words or characters.
During inference, LLM-ForcedAligner enables non-autoregressive decoding, completely avoiding hallucinations compared with autoregressive decoding and achieving faster speed.
Experimental results show that, in multilingual, cross-lingual, and long-form speech scenarios of up to 300 seconds, LLM-ForcedAligner achieves a 69\%\textasciitilde78\% relative reduction in accumulated averaging shift (AAS) compared with prior FA methods, while incurring only a slight increase in real-time factor (RTF).
\begin{figure*}[t]
\centering
\includegraphics[width=1.0\textwidth]{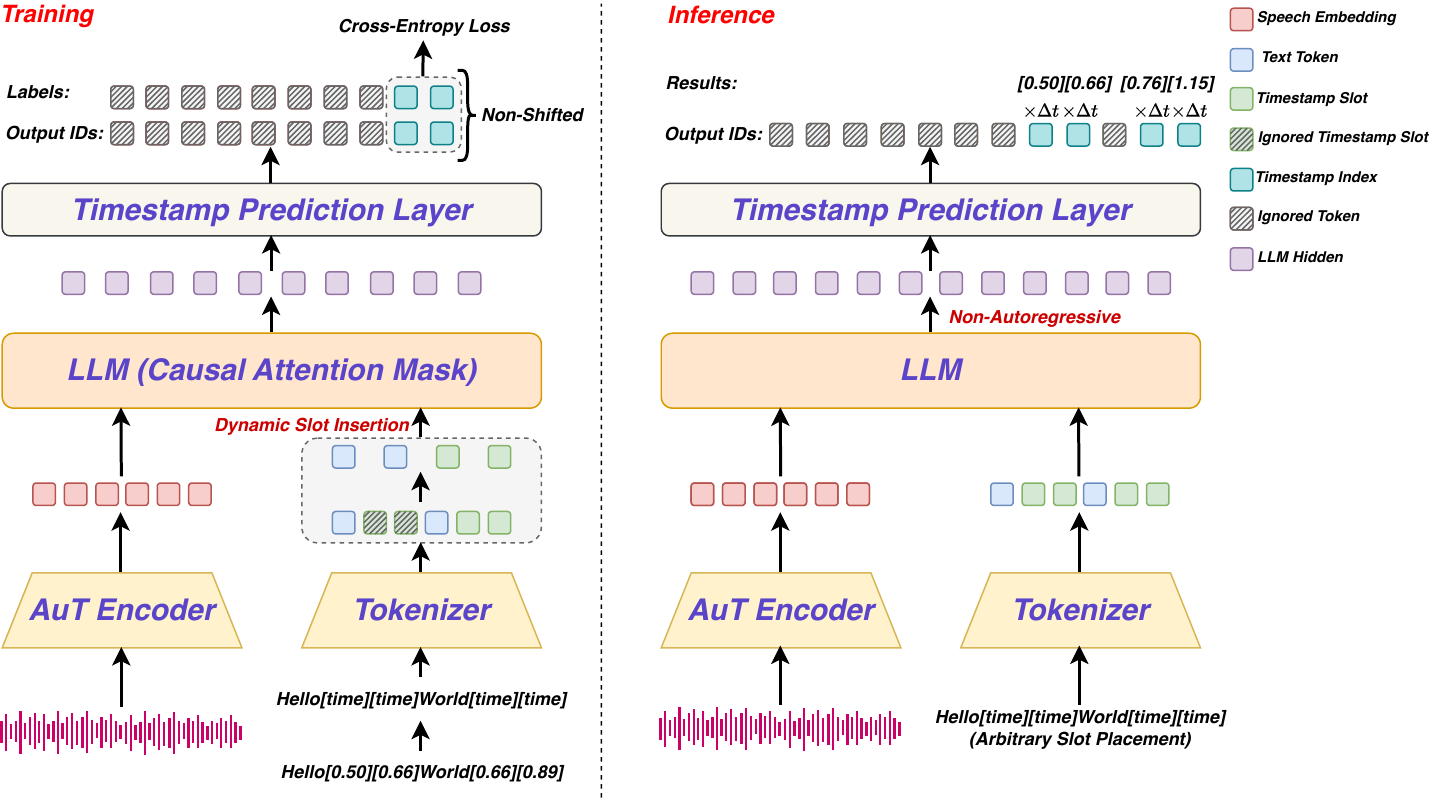} 
\caption{Overview of LLM-ForcedAligner. \textbf{\textit{Left:}} During training, we replace the word-level or character-level start and end timestamps in the transcript with a special token \textit{[time]} to serve as slots, and use dynamic slot insertion to randomly determine which word or character slots to ignore. The speech embeddings produced by the AuT encoder are then concatenated and fed into the LLM for training with causal attention masking. When computing the cross-entropy loss, the output IDs and the labels are non-shifted, and the loss is computed only at the positions of the retained slots. \textbf{\textit{Right:}} During inference, users can insert the special token \textit{[time]} at arbitrary positions in the transcript and rapidly obtain the corresponding start and end timestamps through non-autoregressive decoding.}
\vspace{-11pt}
\label{fig2}
\end{figure*}
\section{Related Work}
\textbf{Traditional Hybrid Systems for FA.} 
These methods~\citep{Michael2017mfa} build GMM–HMM acoustic models based on Kaldi~\citep{povey2011kaldi}, map text to phoneme sequences using lexicons and pronunciation dictionaries, and then apply Viterbi decoding to search for the most probable frame-level alignment path under constrained transitions. Owing to the inherent monotonic temporal structure of HMMs and their fine-grained phoneme modeling, such methods typically achieve high accuracy at phoneme-level boundaries. 
However, they rely heavily on language-specific lexicons, pronunciation dictionaries, and phoneme sets, with each language typically requiring a separately trained or adapted GMM–HMM model.

\noindent\textbf{End-to-End Models for FA.}
These methods no longer rely on explicit lexicon mapping; instead, they use the CTC or internal attention weights to learn a direct mapping between acoustic features and phoneme sequences, achieving alignment by computing emission probabilities. However, they require adaptation for each language and tend to produce systematic temporal shifts.
CTC-based NeMo forced aligner (NFA)~\citep{Elena2023nfa} requires language-specific lexicons, and multilingual CTC models suffer from trade-offs between different languages.
NFA relies on sparse peak frames, which often lag the actual onset of speech, leading to overall backward shifts in token boundaries that accumulate into significant systemic temporal shifts in long-form speech.
CIF~\citep{xian2023achieving} couples with language characteristics, resulting in much lower FA performance in word-boundary languages like English than in character-boundary languages like Chinese.
In CIF, each token is typically accumulated over approximately a fixed number of encoder frames rather than adapting to its actual duration, leading to systematically delayed end times for long vowels, long syllables, or long words. Although post-processing such as scaled-CIF and fire-delay can partially alleviate this issue, such structural shifts remain difficult to eliminate in long-form speech.
WhisperX~\citep{Max2023whisperx} requires switching between phoneme recognition models to accommodate different languages, and remains essentially based on DTW with local similarity. In long-form speech, local alignment errors can similarly accumulate into global temporal shifts.

\noindent\textbf{SLLMs for FA.}
Existing SLLMs~\citep{yunfei2023qwenaudio, xuelong2025osum} treat FA as a by-product of ASR, processing timestamps as natural language, which prevents guaranteed temporal monotonicity during autoregressive inference and can result in excessive timestamps due to LLM hallucinations; in long-form speech scenarios, this significantly increases inference time. Therefore, the next-token prediction paradigm of conventional SLLM training and inference is not well-suited for FA, whereas LLM-ForcedAligner restructures the FA, enabling accurate and fast timestamp prediction.
\section{LLM-ForcedAligner}
\subsection{Overall Architecture}
LLM-ForcedAligner formulates FA as a slot-filling paradigm: given the speech signal and a transcript augmented with special tokens \textit{[time]} that denote word-level or character-level start and end time slots, the SLLM directly predicts the corresponding discrete timestamp indices for each slot.
Unlike previous FA methods, which first perform frame-level or phoneme-level alignment and then aggregate the results into word-level or character-level timestamps, LLM-ForcedAligner directly predicts word-level or character-level timestamp indices.

Training LLM-ForcedAligner requires word-level or character-level timestamp labels for a large number of speech–transcript pairs; however, because manual annotation is prohibitively expensive, we adopt pseudo-timestamp labels generated by MFA, which is the most accurate among existing FA methods. 
It is important to emphasize that MFA pseudo-labels inherently contain noise and systematic shifts. LLM‑ForcedAligner does not merely replicate the MFA outputs; instead, it distills and smooths these pseudo-labels using the SLLM, achieving more stable, lower-shift timestamp predictions.
Given a speech signal $x$ and corresponding transcript sequence $w=(w_1, \dots,w_N)$, MFA is used to obtain the start and end times of each word or character and incorporate them into the transcript, resulting in a sequence $wt=(w_1,t_{1,\text{start}}^{\text{MFA}},t_{1,\text{end}}^{\text{MFA}}, \dots,w_N, t_{N,\text{start}}^{\text{MFA}},t_{N,\text{end}}^{\text{MFA}})$.

The speech encoder in LLM-ForcedAligner is from the Audio Transformer (AuT), and its output speech embeddings based on speech signal $x$ are sampled at 12.5Hz, meaning that each speech embedding frame corresponds to 80ms of the speech signal~\citep{jin2025qwen3omni}.
Before feeding $wt$ into the tokenizer, we replace all timestamps with the special token \textit{[time]} to represent slots, resulting in the input transcript sequence $ws=(w_1,\textit{[time]},\textit{[time]},\dots,w_N,\textit{[time]},\textit{[time]})$.
Moreover, we discretize the timestamps in the sequence $wt$ into timestamp indices:
\begin{equation}
  \tau_{i,\text{start} }=\left \lfloor \frac{t_{i,\text{start}}^{\text{MFA}}}{\Delta t}  \right \rfloor ,\tau_{i,\text{end} }=\left \lfloor \frac{t_{i,\text{end}}^{\text{MFA}}}{\Delta t}  \right \rfloor,
\end{equation}
where $\Delta t$ is 80ms chosen to align with the frame rate of the AuT encoder, resulting in a transcript sequence with discrete timestamp indices $wl=(w_1, \tau_{1,\text{start}}, \tau_{1,\text{end}}, \dots w_N, \tau_{N,\text{start}}, \tau_{N,\text{end}})$.
The speech embeddings from the AuT encoder and the text embeddings from $ws$ are fed into the LLM, and a linear layer with 3,750 (i.e., 500s/80ms) output classes is used to predict timestamp indices for the entire input sequence.

The multilingual and crosslingual capabilities of LLM-ForcedAligner are jointly provided by the speech encoder and the multilingual LLM. Specifically, the AuT encoder, pre-trained on a large-scale multilingual corpus, generates effective frame-level speech embeddings for multiple languages, while the multilingual LLM handles semantic information across different languages. Moreover, the special token \textit{[time]} and timestamp prediction layer do not rely on language-specific phoneme sets or lexicons. Therefore, LLM-ForcedAligner can process multilingual and crosslingual speech-transcript pairs, overcoming the language-specific limitations of previous FA methods.
\subsection{Training Strategy}
SLLMs commonly adopt a training scheme in which the last token of the output ID sequence and the first token of the label sequence are removed, creating a one-position shift between the two sequences; the cross-entropy loss is then computed, achieving the standard next-token prediction paradigm.
However, this paradigm is not suitable for filling timestamp slots.
Instead, we adopt causal training with the output ID and label sequences non-shifted, enabling the LLM-ForcedAligner to explicitly perceive timestamp slots during training and predict the timestamps to be filled into those slots.
Moreover, causal training enables the LLM-ForcedAligner to incorporate prior context when predicting the timestamp for the current slot, ensuring global consistency in timestamp prediction.

During training, we compute the cross-entropy loss only at timestamp slot positions, thereby focusing the training objective of LLM-ForcedAligner on timestamp slot filling.
Given the joint input sequence $x$ and $ws$, the LLM produces a hidden-state sequence $h$ under causal attention masking. For each timestamp slot position $j \in \mathcal{I}_{ts}$, a discrete timestamp index distribution is obtained from the timestamp prediction layer:
\begin{equation}
    p\left(\hat{\tau}_{j} \mid x, ws\right)=\operatorname{softmax}\left(\operatorname{TPL}\left(h_{j}\right)\right),
\end{equation}
where $\operatorname{TPL}$ denotes the timestamp prediction layer, and position $j$ corresponds to a start or end timestamp slot for an arbitrary word or character. The loss function of LLM-ForcedAligner is defined as:
\begin{equation}
    \mathcal{L} = -\frac{1}{\left | \mathcal{I}_{ts}  \right | }\sum_{j\in \mathcal{I}_{ts}}\log{p(\hat{\tau}_{j}={\tau}_{j}\mid x,ws)},
\end{equation}
where ${\tau}_{j}$ is the discrete timestamp index from $wl$.

Furthermore, consistently inserting start and end timestamp slots for every word or character during training would cause LLM-ForcedAligner to rely excessively on previously predicted timestamps.
We propose a dynamic slot insertion strategy during training to enhance the generalization capability of LLM-ForcedAligner.
Specifically, for each sample, we determine with a 50\% probability whether to apply dynamic slot insertion. When dynamic slot insertion is applied, each word or character in the sample has a 50\% chance of having start and end timestamp slots inserted after it. This strategy continuously varies the set of timestamp slot positions $\mathcal{I}_{ts}$, enabling LLM-ForcedAligner to predict start and end timestamps for words or characters at arbitrary positions.

\subsection{Non-Autoregressive Inference}
As the output ID and label sequences are kept non-shifted during training, LLM-ForcedAligner can predict the timestamp indices for all slots in a transcript simultaneously using a non-autoregressive decoding.
Specifically, for a speech–transcript pair, users can customize start and end timestamp slots after any word or character. Given a user-defined timestamp slot position $j \in \mathcal{I}_{ts}$, LLM-ForcedAligner predicts its timestamp index via non-autoregressive decoding, which is then converted to a millisecond-level timestamp $\hat t_{j}$:
\begin{equation}
    \label{eq4}
    \hat t_{j} = \hat \tau_{j} \cdot \Delta t,
\end{equation}
where $\Delta t$ is 80ms.
\section{Experimental Setup}
\subsection{Dataset}
We conduct our experiments on 56,000 hours of data containing 10 languages: Chinese, English, French, German, Italian, Japanese, Korean, Portuguese, Russian, and Spanish.
These data are drawn from a combination of internal resources and open-source datasets, covering a wide range of scenarios such as read speech, conversational speech, podcasts, and meetings, including Aishell~\citep{hui2017aishell1, jiayu2018aishell2}, WenetSpeech~\citep{binbin2022wenetspeech}, Aidatatang\footnote{\url{http://openslr.magicdatatech.com/62/}}, Magicdata\footnote{\url{https://www.openslr.org/68/}}, KeSpeech~\citep{zhiyuan2021kespeech}, LibriSpeech~\citep{Vassil2015librispeech}, GigaSpeech~\citep{guoguo2021gigaspeech}, LibriTTS~\citep{Heiga2019libritts}, Emilia~\citep{Haorui2024emilia}, and MLS~\citep{Vineel2020mls}.
All training and test datasets are annotated with pseudo-timestamps generated by MFA, and we additionally assess the performance of LLM-ForcedAligner on an internal Chinese test dataset with manually annotated timestamps.
The transcripts in the training dataset are obtained from either manual annotations or ASR model predictions, which enhances the generalization ability of LLM-ForcedAligner to varying transcript quality.
Details of the training and test datasets are provided in the~\ref{appedix:data}.
\begin{table*}[t]
    \caption{AAS (ms) $\downarrow$ of LLM-ForcedAligner and other FA methods on \textbf{MFA-labeled} test datasets. The test dataset for each language consists of the raw speech from both the open-source and internal test datasets of that language.}
    \label{tab:table1}
    \centering
\begin{tabular}{lcccc}
\toprule
\textbf{Language} & \textbf{Monotonic-Aligner} & \textbf{NFA} & \textbf{WhisperX} & \textbf{LLM-ForcedAligner} \\ \midrule
Chinese           & 161.1                      & 109.8        & -                 & \textbf{33.1}              \\
English           & -                          & 107.5        & 92.1              & \textbf{37.5}              \\
French            & -                          & 100.7        & 145.3             & \textbf{41.7}              \\
German            & -                          & 122.7        & 165.1             & \textbf{46.5}              \\
Italian           & -                          & 142.7        & 155.5             & \textbf{75.5}              \\
Japanese          & -                          & -            & -                 & \textbf{42.4}              \\
Korean            & -                          & -            & -                 & \textbf{37.2}              \\
Portuguese        & -                          & -            & -                 & \textbf{38.4}              \\
Russian           & -                          & 200.7        & -                 & \textbf{40.2}              \\
Spanish           & -                          & 124.7        & 108.0             & \textbf{36.8}              \\ \midrule
Avg.              & 161.1                      & 129.8        & 133.2             & \textbf{42.9}              \\ \bottomrule
\end{tabular}
\end{table*}

\begin{table*}[t]
    \caption{AAS (ms) $\downarrow$ of LLM-ForcedAligner and other FA methods on \textbf{MFA-labeled} test datasets. The test dataset for each language consists of concatenated speech up to 300 seconds in duration from the raw speech in each language’s open-source and internal test datasets. ``Mixed-Crosslingual'' is the concatenated speech with a duration of up to 300 seconds from arbitrary languages’ open-source and internal test datasets.}
    \label{tab:table2}
    \centering
\begin{tabular}{lcccc}
\toprule
\textbf{Language}  & \textbf{Monotonic-Aligner} & \textbf{NFA} & \textbf{WhisperX} & \textbf{LLM-ForcedAligner} \\ \midrule
Chinese            & 1742.4                     & 235.0        & -                 & \textbf{36.5}              \\
English            & -                          & 226.7        & 227.2             & \textbf{58.6}              \\
French             & -                          & 230.6        & 2052.2            & \textbf{53.4}              \\
German             & -                          & 220.3        & 993.4             & \textbf{62.4}              \\
Italian            & -                          & 290.5        & 5719.4            & \textbf{81.6}              \\
Japanese           & -                          & -            & -                 & \textbf{81.3}              \\
Korean             & -                          & -            & -                 & \textbf{42.2}              \\
Portuguese         & -                          & -            & -                 & \textbf{50.0}              \\
Russian            & -                          & 283.3        & -                 & \textbf{43.0}              \\
Spanish            & -                          & 240.2        & 4549.9            & \textbf{39.6}              \\
Mixed-Crosslingual & -                          & -            & -                 & \textbf{34.2}              \\ \midrule
Avg.               & 1742.4                     & 246.7        & 2708.4            & \textbf{52.9}              \\ \bottomrule
\end{tabular}
\end{table*}

\subsection{Implementation Details}
The AuT encoder in LLM-ForcedAligner contains 316.42M parameters and is initialized from the AuT encoder of Qwen3-Omni\footnote{\url{https://huggingface.co/Qwen/Qwen3-Omni-30B-A3B-Thinking}}. 
The LLM uses Qwen3-0.6B\footnote{\url{https://huggingface.co/Qwen/Qwen3-0.6B}}, and the timestamp prediction layer is a single linear layer with 3,750 output timestamp classes and contains 3.84M parameters.
During training, the AuT encoder, the LLM, and the timestamp prediction layer are jointly optimized using the Adam optimizer with a warm-up scheduler, which increases the learning rate to a peak of 0.0003 after 1,000 steps.
\subsection{Evaluation Metrics}
We use accumulated averaging shift (AAS) to measure the performance of timestamp prediction~\citep{xian2023achieving}.
Lower AAS indicates better timestamp prediction performance.
Specifically, AAS computes the average shift for each timestamp slot, defined as the mean absolute difference between the predicted timestamps and the ground-truth timestamps across all slots in the test dataset:
\begin{equation}
    AAS=\frac{ {\textstyle \sum_{k=1}^{K}} \left | \hat k_i - k_i \right | }{K},
\end{equation}
where $K$ is the total number of timestamp slots in the test dataset, $\hat k_i$ is computed according to Eq.~\ref{eq4}, and $k_i$ can be timestamps obtained from MFA or manual annotations.

\section{Experimental Results}
\subsection{Main Results}
Table~\ref{tab:table1} and Table~\ref{tab:table2} compares LLM-ForcedAligner with other FA methods on MFA-labeled test datasets across multilingual, crosslingual, and long-form speech scenarios.
Monotonic-Aligner\footnote{\url{https://modelscope.cn/models/iic/speech_timestamp_prediction-v1-16k-offline}} supports only Chinese, while NFA\footnote{\url{https://github.com/NVIDIA-NeMo/NeMo/tree/main/tools/nemo_forced_aligner}} and WhisperX\footnote{\url{https://github.com/m-bain/whisperX}} require switching models for different languages. Details of compared FA methods are provided in the~\ref{appedix:otherfa}.
We observe that LLM-ForcedAligner not only supports multiple languages without requiring model switching, but also achieves a 66\%\textasciitilde73\% relative reduction in average AAS on multilingual raw speech compared with other FA methods.
Moreover, LLM-ForcedAligner achieves extremely low average AAS on multilingual and crosslingual long-form speech, a setting where other FA methods fail to perform effectively.


Table~\ref{tab:table3} compares LLM-ForcedAligner with other FA methods on human-labeled Chinese test datasets, covering noisy, crosslingual, and long-form speech scenarios.
We find that the average AAS of LLM-ForcedAligner on the human-labeled test datasets achieves a 68\%\textasciitilde78\% relative reduction compared with other FA methods, showing that training LLM-ForcedAligner with MFA-labeled data generalizes well to real-world conditions.


\begin{table*}[t]
    \caption{ \textbf{
    human-labeled} test datasets. ``Raw'' is the raw speech in the dataset; ``Raw-Noisy'' is the raw speech with added background noise; ``Mixed-60s'' and ``Mixed-300s'' are concatenations of raw speech into durations with maximum lengths of 60 and 300 seconds; ``Mixed-Crosslingual'' is a concatenation of human-labeled raw speech and MFA-labeled multilingual speech.}
    \label{tab:table3}
    \centering
\begin{tabular}{lccc}
\toprule
\textbf{Type}      & \textbf{Monotonic-Aligner} & \textbf{NFA} & \textbf{LLM-ForcedAligner} \\ \midrule
Raw                & 49.9                       & 88.6         & \textbf{27.8}              \\
Raw-Noisy          & 53.3                       & 89.5         & \textbf{41.8}              \\
Mixed-60s          & 51.1                       & 86.7         & \textbf{25.3}              \\
Mixed-300s         & 410.8                      & 140.0        & \textbf{24.8}              \\
Mixed-Crosslingual & -                          & -            & \textbf{42.5}              \\ \midrule
Avg.               & 141.3                      & 101.2        & \textbf{32.4}              \\ \bottomrule
\end{tabular}
\end{table*}

\begin{table}[]
    \caption{The average RTF $\downarrow$ of LLM-ForcedAligner and other compared FA methods during inference.}
    \label{tab:table4}
    \centering
\begin{tabular}{lc}
\toprule
\textbf{Methods}       & \textbf{Avg. RTF} \\ \midrule
Monotonic-Aligner      & 0.0079            \\
NFA                    & 0.0067            \\
WhisperX               & 0.0113            \\ 
LLM-ForcedAligner & 0.0159            \\ \bottomrule
\end{tabular}
\end{table}

In addition, in Table~\ref{tab:table2}, the average AAS for long-form speech on the MFA-labeled test datasets is slightly higher than that for raw speech, reflecting the systematic shifts of MFA on long-form speech. In contrast, in Table~\ref{tab:table3}, on the human-labeled test datasets, the average AAS for long-form speech is lower than that for raw speech, indicating that LLM-ForcedAligner does not simply replicate MFA’s timestamp predictions, but instead learns a more robust and reliable timestamp prediction that can effectively correct MFA labels in long-form scenarios. Furthermore, during long-form inference, LLM-ForcedAligner can leverage a longer historical context to predict timestamps for the current slots, resulting in superior performance on human-labeled long-form speech test datasets.

Table~\ref{tab:table4} reports the average RTF of LLM-ForcedAligner and other compared FA methods under identical inference conditions.
As the number of model parameters increases, the RTF shows a slight increase.
Due to the benefit of the non-autoregressive inference of LLM-ForcedAligner, it achieves a substantial reduction in average AAS with only a minimal increase in RTF.
Users can select the most suitable FA method based on the average AAS-RTF trade-off.

\subsection{Ablation Study}
Table~\ref{tab:table5} shows the average AAS results of LLM-ForcedAligner trained with different timestamp token durations on the MFA-labeled and human-labeled test datasets.
When the timestamp token duration is 120ms, the timestamp prediction layer has 2,500 classes (i.e., 300s/120ms); when the duration is 80ms, it has 3,750 classes (i.e., 300s/80ms); and when the duration is 40ms, it has 7,500 classes (i.e., 300s/40ms).
As the timestamp token duration decreases, the AAS on the MFA-labeled test datasets steadily declines, indicating that finer-grained timestamp prediction better fits the MFA labels.
However, on human-labeled test datasets, finer-grained timestamp prediction does not yield lower AAS because it better fits the MFA timestamp distribution, leading to reduced generalization.
The timestamp token duration of 80ms is the optimal choice, as each frame of the AuT encoder’s output also represents 80ms of speech, which helps LLM-ForcedAligner better determine the start and end timestamps of words or characters based on speech boundaries.

\begin{table}[]
    \caption{AAS (ms) $\downarrow$ on \textbf{MFA-labeled} and \textbf{human-labeled} test datasets for the ablation study of different timestamp token durations.. ``Raw'' is raw speech, while ``Mixed'' is monolingual and crosslingual concatenated speech up to 300 seconds.}
    \label{tab:table5}
    \centering
\scalebox{0.86}{
\begin{tabular}{lcccc}
\toprule
\multirow{2}{*}{\textbf{Timestamp Dur.}} & \multicolumn{2}{c}{\textbf{MFA-Labeled}} & \multicolumn{2}{c}{\textbf{Human-Labeled}} \\ \cmidrule{2-5} 
                                               & \textbf{Raw}       & \textbf{Mixed}      & \textbf{Raw}       & \textbf{Mixed}       \\ \midrule
120ms                                          & 51.1               & 62.9                & 35.5               & 34.5                 \\
80ms                                           & 41.7               & 52.9                & \textbf{27.8}               & \textbf{25.1}                 \\
40ms                                           & \textbf{34.0}               & \textbf{50.9}                & 32.5               & 27.9                 \\ \bottomrule
\end{tabular}}
\end{table}

Table~\ref{tab:table6} shows the average AAS results of LLM-ForcedAligner on the MFA-labeled and human-labeled test datasets, comparing results with and without dynamic slot insertion during training.
Dynamic slot insertion randomly determines whether to insert timestamp slots after each word or character, enabling LLM-ForcedAligner to predict start and end timestamps at arbitrary positions and preventing it from relying excessively on previously predicted timestamps.
We find that dynamic slot insertion reduces AAS on both test datasets, with the improvement more pronounced for long-form speech. This phenomenon is because dynamic slot insertion, by randomly deciding whether to insert timestamp slots after each word or character, prevents LLM-ForcedAligner from excessively relying on historically predicted timestamps, which can otherwise lead to systematic temporal shifts. Furthermore, dynamic slot insertion enables LLM-ForcedAligner to predict start and end timestamps for words or characters at arbitrary positions, supporting user-customizable timestamp prediction.
\begin{table}[]
    \caption{AAS (ms) $\downarrow$ of the ablation study on dynamic slot insertion across \textbf{MFA-labeled} and \textbf{human-labeled} test datasets. ``Raw'' is raw speech, while ``Mixed'' is monolingual and crosslingual concatenated speech up to 300 seconds.}
    \label{tab:table6}
    \centering
\scalebox{0.76}{
\begin{tabular}{lcccc}
\toprule
\multirow{2}{*}{\textbf{Dynamic Slot Insertion}} & \multicolumn{2}{c}{\textbf{MFA-Labeled}} & \multicolumn{2}{c}{\textbf{Human-Labeled}} \\ \cmidrule{2-5} 
                                                 & \textbf{Raw}       & \textbf{Mixed}      & \textbf{Raw}       & \textbf{Mixed}       \\ \midrule
w/o                                              & 51.2               & 66.1                & 31.4               & 30.4                 \\
w/                                               & \textbf{41.7}               & \textbf{52.9}                & \textbf{27.8}               & \textbf{25.1}                 \\ \bottomrule
\end{tabular}}
\end{table}

\begin{figure}[t]
\centering
\includegraphics[width=1.0\columnwidth]{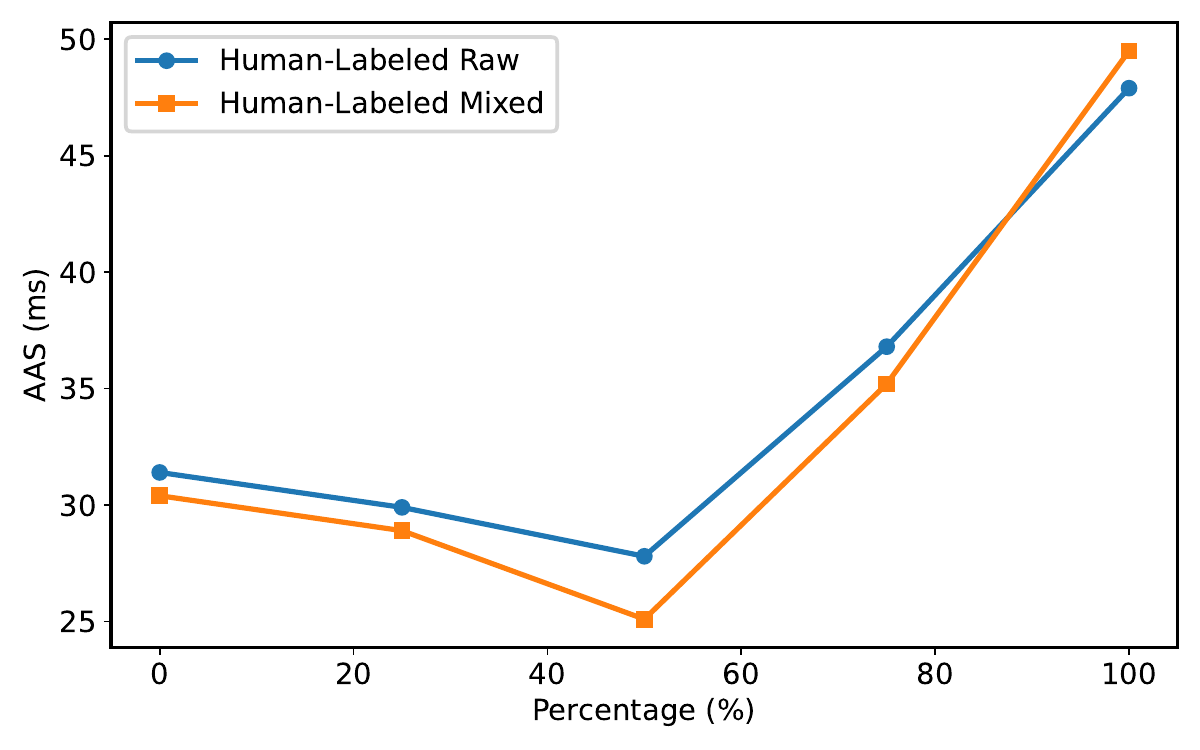}
\caption{AAS (ms) of LLM-ForcedAligner on the \textbf{human-labeled} test datasets for different dynamic slot insertion percentages.}
\label{fig3}
\end{figure}

\subsection{Visualization}
Figure~\ref{fig3} shows the AAS results on the human-labeled test datasets when LLM-ForcedAligner is trained with different percentages of dynamic slot insertion.
When the percentage of dynamic slot insertion is below 50\% of the training samples, LLM-ForcedAligner achieves lower AAS, and the AAS continues to decrease as the percentage increases. However, when the percentage exceeds 50\% of the training samples, the AAS begins to increase, reaching its highest value at 100\%. Therefore, selecting an appropriate dynamic slot insertion percentage of 50\% is crucial for enhancing the generalization of LLM-ForcedAligner.

\begin{figure}[t]
\centering
\includegraphics[width=1.0\columnwidth]{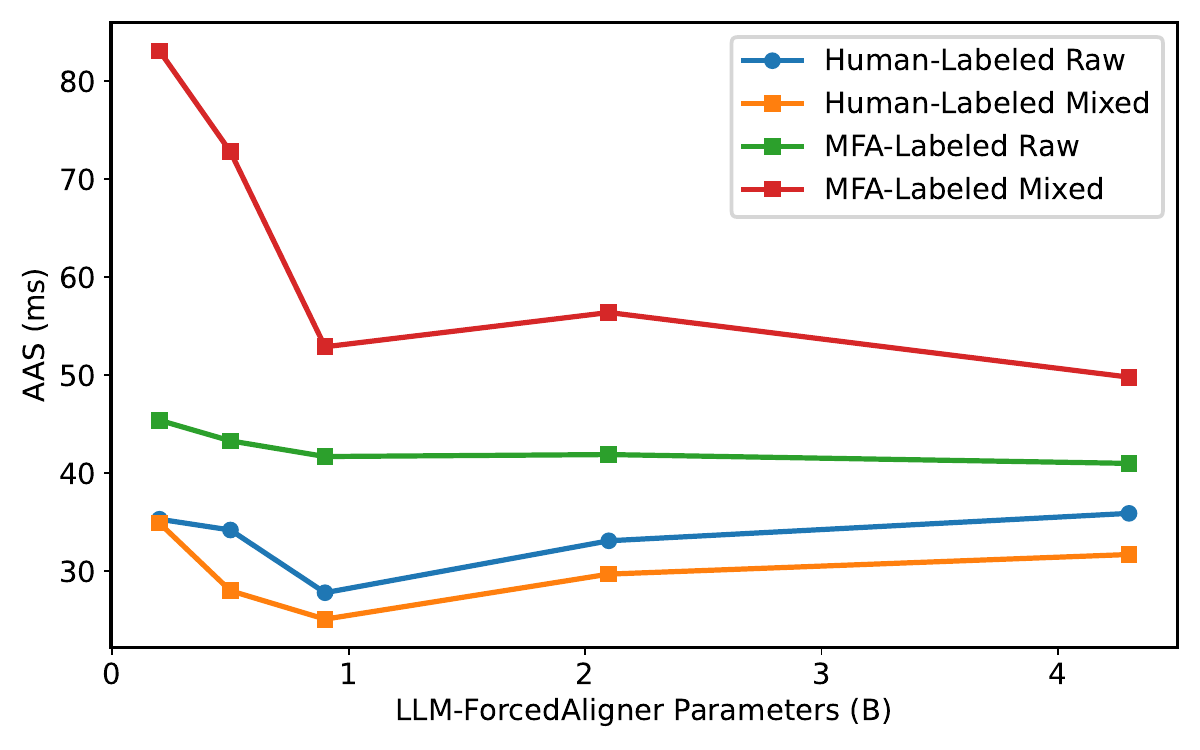}
\caption{AAS (ms) on the \textbf{MFA-labeled} and \textbf{human-labeled} test datasets for LLM-ForcedAligner with different parameter settings.}
\label{fig4}
\end{figure}

Figure~\ref{fig4} shows the AAS results on the MFA-labeled and human-labeled test datasets for LLM-ForcedAligner with different parameter settings.
When the LLM-ForcedAligner parameter size is below 0.9B, timestamp prediction performance is limited by insufficient model capacity. 
When the parameter size exceeds 0.9B, the AAS on the MFA-labeled test datasets shows no significant change, while the AAS on the human-labeled test datasets increases, indicating that LLM-ForcedAligner overfits the MFA timestamp distribution. 
A parameter size of 0.9B is therefore optimal. At this scale, LLM-ForcedAligner does not strictly fit the MFA timestamp distribution, but instead learns a smoother and more robust timestamp prediction behavior with better generalization performance.
\section{Conclusion}
We propose \textbf{LLM-ForcedAligner}, a reformulation of FA as a slot-filling paradigm: timestamps are treated as discrete indices, and special timestamp tokens are inserted as slots into the transcript. Conditioned on the speech and the transcript with slots, LLM-ForcedAligner directly predicts the time indices at slots. During training, causal attention masking with non-shifted input and label sequences allows each slot to predict its own timestamp index based on itself and prior context, with loss computed only at slot positions. Dynamic slot insertion enables timestamp prediction at arbitrary positions. Non-autoregressive inference is supported, avoiding hallucinations and improving speed. Experiments show that LLM-ForcedAligner is an accurate, fast and customized FA method for multilingual, crosslingual, and long-form speech scenarios.
\clearpage
\section*{Limitations}
Manually annotated millisecond-level timestamps also suffer from ambiguous temporal boundaries, making it difficult for any method to achieve highly precise alignment. Therefore, we train LLM-ForcedAligner on MFA-labeled pseudo-timestamp datasets and evaluate it on manually annotated timestamp datasets, thereby assessing whether the predicted timestamps exhibit perceptible deviations from human judgments.
We generate pseudo-timestamp labels using the MFA method to construct most of the training and test datasets, which causes LLM-ForcedAligner to fit the relatively accurate MFA timestamp-prediction distribution.
Although the manually annotated test dataset allows us to assess LLM-ForcedAligner’s performance in practical scenarios, it covers only Chinese and cannot verify the actual performance in other languages. 
In addition, the training dataset exhibits an uneven language distribution, which may explain why the AAS for other languages is higher than that for Chinese and English. In future work, we plan to explore methods to improve LLM-ForcedAligner’s performance across other languages efficiently and to extend its application scenarios to more challenging conditions, such as meetings, music, and film or television content.

\bibliography{custom}
\appendix

\section{Appendix}
\subsection{Data Statistics}\label{appedix:data}
\begin{table}[h]
    \caption{Overall dataset statistics.}
    \label{tab:table7}
    \centering
\scalebox{0.82}{
\begin{tabular}{lcc}
\toprule
\textbf{Language} & \textbf{Sources}                                                                                                                & \textbf{Hours} \\ \midrule
Chinese           & \begin{tabular}[c]{@{}c@{}}AISHELL-1, AISHELL-2, \\ WenetSpeech, Aidatatang\_200zh, \\ Magicdata, KeSpeech, Private data\end{tabular} & 31536          \\
English           & \begin{tabular}[c]{@{}c@{}}LibriSpeech, GigaSpeech, \\ VCTK, LibriTTS, Private data\end{tabular}                                    & 20322          \\
French            & Emilia, MLS                                                                                                                     & 611            \\
German            & Emilia, MLS                                                                                                                     & 517            \\
Italian           & MLS, Private data                                                                                                                   & 602            \\
Japanese          & Emilia                                                                                                                          & 731            \\
Korean            & Emilia, Private data                                                                                                                & 662            \\
Portuguese        & MLS, Private data                                                                                                                   & 316            \\
Russian           & Private data                                                                                                                        & 268            \\
Spanish           & MLS, Private data                                                                                                                   & 579            \\ \bottomrule
\end{tabular}}
\end{table}
Table~\ref{tab:table7} summarizes the sources and durations of the overall datasets used for LLM-ForcedAligner. Consequently, we provide a brief introduction to the open-source datasets included.

\noindent\textbf{AISHELL-1}~\citep{hui2017aishell1}: The dataset contains 400 speakers and over 170 hours of Mandarin speech data, covering 5 topics: ``Finance'', ``Science and Technology'', ``Sports'', ``Entertainments'', and ``News''. Transcripts are manually filtered to eliminate improper contents.

\noindent\textbf{AISHELL-2}~\citep{jiayu2018aishell2}: The dataset contains 1991 speakers and over 1000 hours of clean reading speech data. The content of the recording covers 8 major topics: voice commands such as IoT device control and digital sequential input, places of interest, entertainment, finance, technology, sports, English spellings and free speaking without specific topic. 

\noindent\textbf{WenetSpeech}~\citep{binbin2022wenetspeech}: The dataset is a multi-domain Mandarin corpus consisting of 10000+ hours high-quality labeled speech, 2400+ hours weakly labeled speech, and about 10000 hours unlabeled speech, with 22400+ hours in total. The data from YouTube and Podcast, which covers a variety of speaking styles, scenarios, domains, topics and noisy conditions. 

\begin{table*}[]
    \caption{Configuration of Compared FA methods. ``-'' indicates that no model is available for this language, all model names are identical to the corresponding official open-source model names.}
    \label{tab:table8}
    \centering
\scalebox{0.85}{
\begin{tabular}{lccc}
\toprule
\textbf{Language} & \textbf{Monotonic-Aligner} & \textbf{NFA}                              & \textbf{WhisperX}             \\ \midrule
Chinese           & Paraformer                 & stt\_zh\_citrinet\_1024\_gamma\_0\_25     & -                             \\
English           & -                          & stt\_en\_fastconformer\_hybrid\_large\_pc & WAV2VEC2\_ASR\_BASE\_960H     \\
French            & -                          & stt\_fr\_conformer\_ctc\_large            & VOXPOPULI\_ASR\_BASE\_10K\_FR \\
German            & -                          & stt\_de\_fastconformer\_hybrid\_large\_pc & VOXPOPULI\_ASR\_BASE\_10K\_DE \\
Italian           & -                          & stt\_it\_fastconformer\_hybrid\_large\_pc & VOXPOPULI\_ASR\_BASE\_10K\_IT \\
Japanese          & -                          & -                                         & -                             \\
Korean            & -                          & -                                         & -                             \\
Portuguese        & -                          & -                                         & -                             \\
Russian           & -                          & stt\_ru\_quartznet15x5                    & -                             \\
Spanish           & -                          & stt\_es\_fastconformer\_hybrid\_large\_pc & VOXPOPULI\_ASR\_BASE\_10K\_ES \\ \bottomrule
\end{tabular}}
\end{table*}

\noindent\textbf{Aidatatang\_200zh}\footnote{\url{http://openslr.magicdatatech.com/62/}}: The dataset is a Chinese Mandarin speech corpus, containing 200 hours of speech data from 600 speakers. The transcription accuracy for each sentence is larger than 98\%.

\noindent\textbf{Magicdata}\footnote{\url{https://www.openslr.org/68/}}: The dataset contains 755 hours of scripted read speech data from 1080 native speakers of the Mandarin Chinese spoken in mainland China. The sentence transcription accuracy is higher than 98\%. The domain of recording texts is diversified, including interactive Q\&A, music search, SNS messages, home command and control, etc.

\noindent\textbf{KeSpeech}~\citep{zhiyuan2021kespeech}: The dataset comprises 1,542 hours of speech signals recorded by 27,237 speakers across 34 cities in China, with pronunciation in standard Mandarin and 8 subdialects. Two professional data companies manually label the dataset with three steps.

\noindent\textbf{LibriSpeech}~\citep{Vassil2015librispeech}: The dataset is derived from audiobooks that are part of the LibriVox project, contains 1000 hours of speech sampled at 16kHz, and has been carefully segmented and aligned.

\noindent\textbf{GigaSpeech}~\citep{guoguo2021gigaspeech}: The dataset is a multi-domain English speech recognition corpus with 10,000 hours of high quality labeled audio suitable for supervised training, and 40,000 hours of total audio suitable for semi-supervised and unsupervised training.

\noindent\textbf{LibriTTS}~\citep{Heiga2019libritts}: The dataset is a multi-speaker English corpus consisting of approximately 585 hours of read English speech at a 24kHz sampling rate from 2,456 speakers, along with the corresponding texts.

\noindent\textbf{MLS}~\citep{Vineel2020mls}: The dataset is derived from read audiobooks from LibriVox and consists of 8 languages, including about 44.5K hours of English and about 6K hours of other languages.

\noindent\textbf{Emilia}~\citep{Haorui2024emilia}: The dataset contains over 101k hours of speech data at 24 kHz and covers six languages. It comprises mostly spontaneous
speech, covering a wide range of speaking styles.

\noindent\textbf{Private Data}: The datasets are primarily sourced from professional data providers and encompass multilingual read and conversational speech, and all annotations are manually reviewed.

\noindent\textbf{Construction of Training Dataset}: All data in Table~\ref{tab:table7} are first annotated with word-level or character-level start and end timestamp pseudo-labels obtained using MFA. For each language and each data source, a small portion is reserved as the test dataset. From the remaining data, 30\% is used as raw speech for training, while the remaining 70\% is used to construct long-form speech through mixing. Specifically, a target duration is randomly sampled from a uniform distribution between 30s and 500s, and raw speech from different languages is randomly concatenated until the target duration is reached. To further enhance data diversity, various types of noise are randomly mixed into the mixed speech.

\noindent\textbf{Construction of Test Datasets}: For each language and data source, 30\% of the reserved raw speech is used as the MFA-labeled Raw test datasets, while the remaining 70\% is used to create long-form speech through mixing. The mixing procedure is the same as for the training dataset, but divided into monolingual and crosslingual mixing. The resulting mixed test datasets serve as the MFA-labeled Mixed test datasets. In addition, we evaluate the performance of LLM-ForcedAligner on an internal, manually annotated Chinese data source. Specifically, this source is first used as the human-labeled Raw test dataset, and background noise is then added to create the human-labeled Raw-Noisy test dataset. It is also concatenated to form the human-labeled Mixed-60s and human-labeled Mixed-300s test datasets, each with a maximum duration of 60s and 300s, respectively. Finally, it is combined with the MFA-labeled Raw test dataset to create the human-labeled Mixed-Crosslingual test dataset.

\subsection{Details of Compared FA Methods}\label{appedix:otherfa}
Since the compared FA methods require switching backbone models across languages, Table~\ref{tab:table8} lists the backbone model used by each method for each language in the comparison experiments. We strictly reproduce the results on our test datasets following the standard evaluation procedures of these FA methods.

\noindent\textbf{Monotonic-Aligner}~\citep{xian2023achieving}: It supports only Chinese and uses Paraformer as its backbone model. Its standard evaluation procedures are on its homepage\footnote{\url{https://modelscope.cn/models/iic/speech_timestamp_prediction-v1-16k-offline}}.

\noindent\textbf{NFA}~\citep{Elena2023nfa}: It is a tool for generating token-, word-, and segment-level timestamps of speech in audio using NeMo's CTC-based ASR models.
Its standard evaluation procedures are on its homepage\footnote{\url{https://github.com/NVIDIA-NeMo/NeMo/tree/main/tools/nemo_forced_aligner}}, and the available checkpoint list is on the NeMo ASR collection page\footnote{\url{https://catalog.ngc.nvidia.com/orgs/nvidia/collections/nemo_asr}}.

\noindent\textbf{WhisperX}~\citep{Max2023whisperx}: It performs FA using phoneme recognition models specific to different languages. Its standard evaluation procedures are on its homepage\footnote{\url{https://github.com/m-bain/whisperX}}, and the available checkpoint list is on the inference script file\footnote{\url{https://github.com/m-bain/whisperX/blob/main/whisperx/alignment.py}}.

\end{document}